\documentclass{article}

\usepackage{arxiv}

\usepackage{float}
\usepackage[utf8]{inputenc}
\usepackage[T1]{fontenc}   
\usepackage{hyperref}   
\usepackage{url}      
\usepackage{booktabs}   
\usepackage{amsfonts}
\usepackage{nicefrac}   
\usepackage{microtype} 
\usepackage{lipsum}
\usepackage{amsmath}
\usepackage{graphicx}
\usepackage{subfig}

\newcommand{\pass}{\ensuremath{\textit{pass}}}
\newcommand{\avgnist}{\ensuremath{\text{avg}_\text{NIST}}}

\title{Pseudo Random Number Generation: a Reinforcement Learning approach}

\author{
  Luca Pasqualini\thanks{Corresponding author: http://sailab.diism.unisi.it/people/luca-pasqualini/} \\
  Department of Information Engineering and Mathematical Sciences\\
  University of Siena\\
  Siena, 53100 Italy \\
  \texttt{pasqualini@diism.unisi.it}
  \And
  Maurizio Parton\\
  Department of Economical Studies\\
  University of Chieti-Pescara\\
  Pescara, 65129 Italy\\
  \texttt{parton@unich.it}\\
}

\begin{document}
\maketitle
\begin{abstract}
  Pseudo-Random Numbers Generators (PRNGs) are algorithms produced to generate long sequences of statistically uncorrelated numbers, i.e.\ Pseudo-Random Numbers (PRNs). These numbers are widely employed in mid-level cryptography and in software applications. Test suites are used to evaluate PRNGs quality by checking statistical properties of the generated sequences. 
  Machine learning techniques are often used to break these generators, for instance approximating a certain generator or a certain sequence using a neural network. But what about using machine learning to generate PRNs generators?
  This paper proposes a Reinforcement Learning (RL) approach to the task of generating PRNGs from scratch by learning a policy to solve an $N$-dimensional navigation problem. In this context, $N$ is the length of the period of the generated sequence, and the policy is iteratively improved using the average value of an appropriate test suite run over that period.
  Aim of this work is to demonstrate the feasibility of the proposed approach, to compare it with classical methods, and to lay the foundation of a research path which combines RL and PRNGs.
\end{abstract}

\section{Introduction}
Pseudo-Random Numbers (PRNs) sequence generation is a task of renown importance in cryptography and more in general in computer science. A Pseudo-Random Number Generator (PRNG) is a (usually, deterministic) algorithm which tries to emulate the statistical properties of a sequence of True-Random Numbers (TRNs). PRNGs are used in applications related to gambling, statistical sampling, computer simulation and in other areas where producing an unpredictable result is desirable. 
While TRN sequences are overall more unpredictable and as such better keys for cryptography systems, they are usually expensive to generate. Indeed, a True Random Number Generator (TRNG) relies on natural phenomena like atmospheric or thermal noise, radioactive decay or cosmic background radiation. Measurement of these is known to be expensive.
As a consequence TRNGs are employed only in applications of vital importance while PRNGs are used in almost every other setting. 

A wide variety of methods have been employed up to now to build PRNGs. One of the earlier type of generators was based on linear recurrences. Generators belonging to this family are linear congruential generators, linear feedback registers and xorshifts generators, to name a few. Today, cryptographically secure PRNGs are designed and employed by major industries. Some example are stream ciphers, block ciphers or combination generators which combines multiple primitive PRNGs to improve the quality of the result.

To measure the quality of a PRNG some kind of statistical test is run over the result, i.e.\ the generated sequence. Usually, more tests are combined in what is called a test suite. The overall accuracy of a test suite usually increases with the length of the sequence generated by the algorithm. Moreover, the sequence is usually analyzed in its binary format. This is why PRNGs are also called Deterministic Random Bit Generators (DRBGs). They actually generate sequences of uncorrelated bits which can be interpreted as integers or floating point numbers depending on the chosen representation. In this paper the National Institute for Standard and Technologies (NIST) statistical test suite for random and pseudo-random number generators \cite{bassham2010sp} is used to train and validate the PRNG.

Machine Learning (ML) is a field of artificial intelligence studying algorithms and statistical models to be used by computer systems to perform tasks without explicit instructions. Nowadays, widely used statistical models are a class of function approximators called Neural Networks (NNs). In particular, Deep Neural Networks (DNNs), that is, NNs with several hidden layers, are successfully employed in many fields like image recognition, natural language processing, games, etc.

For the above reasons, ML has been used in the field of PRNGs to approximate generators using target sequences of pseudo-random or true-random bits. This technique is very useful when the goal is predicting the output of an existing generator, e.g.\ to break the key of a cryptography system.
There have been limited attempts at generating PRNGs using NNs by exploiting their structure and internal dynamics. For example, the authors of \cite{desai2012using} use Recurrent Neural Networks (RNNs) dynamics to generate PRNs. Additional approaches of this type has been investigated since NNs are highly non-linear mathematical systems. In \cite{hughes2007pseudo}, the authors use the dynamics of a feed forward NN with random orthogonal weight matrices to generate PRNs. Neuronal plasticity is used in \cite{abdi1994neural} instead. In \cite{marcellopseudo} a Generative Adversarial Network (GAN) approach to the task is presented, exploiting an input source of randomness (like an existing PRNG or a TRNG) and the dynamics of two NNs acting in adversarial way.

This paper presents a novel approach to the task of generating PRNGs from scratch. The proposed approach works without data nor external inputs and without employing any structural dynamics. Indeed, it works just by using Deep Reinforcement Learning (DRL), that is, RL and a DNN as function approximator.
Our approach generates a sequence of PRNs with variable period by directly optimizing its value, as computed by the NIST test suite. We propose a decimal and a binary formulation, and assess how each formulation performs with respect to different RL algorithms.

The paper is organized as follows: in section two we model the task of generating PRNs as a Markov Decision Process (MDP) in order to tackle it with RL techniques.
We also cover in more detail what RL is and how the NIST value used to train the RL algorithms is computed. Section three is where experimental setup and results are described. Section four is where conclusions are drawn and future work is proposed. In particular, in section four we show both advantages and current limits of this RL approach, and plan a strategy to overcome them with future research.

\section{Methodology}
When a PRNG generates a sequence of bits it manipulates a starting state, called seed, according to a certain algorithm that generates the next state, and so on. It can be thought as a sequence of decisions made from a starting state following a certain strategy. In this setting, a good PRNG corresponds to a good strategy in a decision-making process. To put in practice this raw considerations, we formulate the problem as a RL task.

\subsection{Reinforcement Learning}

For a comprehensive, motivational and thorough introduction to RL, we strongly suggest reading sections from $1.1$ to $1.6$ in \cite{sutton2018reinforcement}. In what follows, we summarize the terminology used in this paper.

First of all, RL is learning what to do in order to accumulate as much reinforcement as possible during the course of our actions. This very general description, known as \emph{the RL problem}, can be framed as a sequential decision-making problem as follows.

Assume an \emph{agent} is interacting with an \emph{environment}. When the agent is in a certain situation - a \emph{state} - it has several options - called \emph{actions}. After each action, the environment will take the agent to a next state, and will provide it with a numerical \emph{reward}, where the pair "state, reward" may possibly be drawn from a joint probability distribution, called the \emph{model} of the environment. The agent will choose actions according to a certain strategy, called \emph{policy} in the RL setting. The RL problem can then be stated as finding a policy maximizing the expected value of the total reward accumulated during the interaction agent-environment.

In order to formalize the above description, we start from the classical picture representing the agent-environment interaction.
	\begin{figure}[!tbp]
	\centering
	\includegraphics[width=0.75\textwidth]{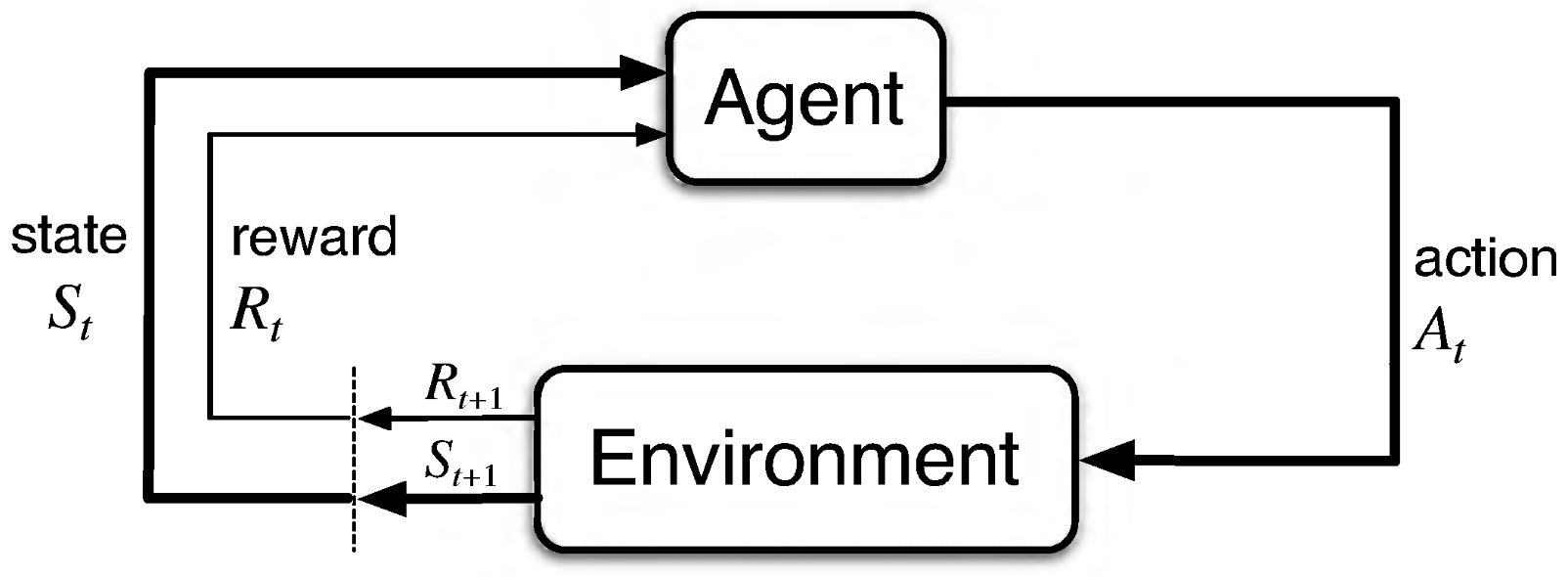}
	\caption{\cite[figure 3.1]{sutton2018reinforcement} The agent-environment interaction is made at discrete time steps $t=0,1,2,\dots$}
		\label{fig:mdp}
	\end{figure}
At each time step $t$, the agent receives a state $S_t\in\mathcal{S}$ from the environment, and then selects an action $A_t\in\mathcal{A}$. The environment answers with a numerical reward $R_{t+1}\in\mathcal{R}\subset\mathbb{R}$ and a next state $S_{t+1}$. This interaction gives raise to a \emph{trajectory} of random variables:
\[
S_0,A_0,R_1,S_1,A_1,R_2,\dots
\]
In the case of interest to us, $\mathcal{S},\mathcal{A}$ and $\mathcal{R}$ are finite sets. Thus, the environment answers the action $A_t=a$ executed in the state $S_t=s$ with a pair $R_{t+1}=r,S_{t+1}=s'$ drawn from a discrete probability distribution $p$ on $\mathcal{S}\times\mathcal{R}$, known as the model or the \emph{dynamics} of the environment:
\[
p(s',r,s,a):=\text{Pr}(S_{t+1}=s',R_{t+1}=r|S_t=s,A_t=a).
\]
To give a visual clue to the fact that the $p(s',r,s,a)$ is a probability distribution when $s,a$ are fixed, it is customary to write $p(s',r|s,a)$ instead.
It is worth noticing that we are implicitly making two assumptions: the model $p$ is stationary, that is, $p$ does not change with time (otherwise, we should write $p_t$ instead of $p$), and the joint probability distribution of $S_{t+1},R_{t+1}$ depends only on the previous state $S_t$ and action $A_t$, and nothing else from the trajectory (in figure~\ref{fig:mdp} the environment is fed only with last action, and no other data from the history). This last property, resounding the Markov property of a stochastic process, gives the name \emph{Markov Decision Process} (MDP) to the data $(\mathcal{S},\mathcal{A},\mathcal{R},p)$. Note that, more than a restriction of the decision process, the Markov property should be thought as a restriction on the state: it must include all aspects of the past that are needed for the future.

How does an agent learn in this finite MDP setting? When the agent experiences a trajectory starting at time $t$, it accumulates a \emph{discounted return}\footnote{A discount factor $\gamma<1$ is used mainly in two classes of problems: in \emph{continuing} tasks, that is, when the trajectories do not decompose naturally into \emph{episodes}, it is needed to make finite the infinite sum; and when rewards far in the future are less and less reliable or important. Usually, $\gamma$ is given as data of the MDP definition, but it seems more natural to consider it part of the learning task, since $\gamma$ is used by the agent to choose his actions.} $G_t$:
\[
G_t:=R_{t+1}+\gamma R_{t+2}+\gamma^2 R_{t+3}+\dots=\sum_{k=0}^\infty\gamma^k R_{t+k+1},\qquad\gamma\in[0,1].
\]
It should be now apparent that $G_t$ is a random variable, whose probability distribution depends not only on the model $p$, but also on how the agent chooses actions in a certain state $s$. This is encoded by a discrete probability distribution $\pi$ (the policy) on $\mathcal{A}$:
\[
\pi(a|s):=\pi(a,s):=\text{Pr}(A_t=a|S_t=s).
\]
The total reward the agent can accumulate starting from state $s$, that is, how good is the state $s$ for an agent following the policy $\pi$, is then encoded by the \emph{state-value} function:
\[
v_{\pi}(s):=\mathbb{E}_\pi[G_t|S_t=s].
\]
The RL problem can now be stated as finding a policy $\pi_*$ such that, for every policy $\pi$ and for every state $s$ one has $v_*(s):=v_{\pi_*}(s)\ge v_\pi(s)$. Every policy with this property is called \emph{optimal}, and learning can be tought as the process of finding or approximating an optimal policy.

Another important quantity associated to a policy $\pi$ is the \emph{action-value} function, encoding how good is choosing an action $a$ from a state $s$ and then following the policy $\pi$:
\[
q_{\pi}(s,a):=\mathbb{E}_\pi[G_t|S_t=s,A_t=a].
\]
This is known also as the \emph{quality} or the \emph{q-value} of the action $a$ from the state $s$. 
A policy $\pi_*$ is then optimal if and only if for every state-action pair $s,a$ one has $q_*(s,a):=q_{\pi_*}(s,a)=\max_\pi q_\pi(s,a)$.

How can we learn? That is, how can we find an optimal policy? The starting point is a simple remark: the return $G_t$ is recursively given by $R_{t+1}+\gamma G_{t+1}$. This implies that the RL problem has optimal substructure, expressed as recursive equations for $v_*$ and $q_*$, and therefore dynamic programming can be applied. These recursive equations are known as \emph{Bellman optimality equations}. 

When an accurate model is available, one can use dynamic programming iterative techniques to provide an approximate solution to the Belmann optimality equations. Then, from $v_*$ or $q_*$ one can easily recover an optimal policy: for instance, $\pi_*(s):=\text{argmax}_{a\in\mathcal{A}}q_(s,a)$ is a deterministic optimal policy. 

However, a model of the environment is not always available or feasible to use. In these cases, we must resort to iterative techniques avoiding the model entirely by estimating the value function (\emph{model-free} RL), or to techniques that estimate and use both the value function and the model (\emph{model-based} RL). All approaches presented in this paper belong to model-free RL.

When estimating the action-value function, it is mandatory to sample every pair $s,a$: if from a certain state $s$ the action $a$ was never tried, how can we say whether $a$ is optimal from $s$? Thus, we need to \emph{explore} every action from a given state. On the other hand, we know that the optimal policy we are looking for is essentially deterministic (can be stochastic only on optimal actions), that is, $\pi_*$ \emph{exploits} the best action from every state. This is called the \emph{exploration-exploitation dilemma}, and it can be tackled in two very different ways.

The first approach is considering only soft policies, that is, policies with positive probability for every pair $s,a$. While learning, we give less and less probability to suboptimal actions. This approach is called \emph{on-policy} RL, because we learn the same policy used to explore.
The second approach, called \emph{off-policy} RL, uses two different policies: the behavior policy, used to explore, and the target policy, improved using the returns sampled by the behavior policy. In this way, the target policy can be deterministic, without affecting the exploration.
In this paper, algorithms following both on-policy and off-policy approaches are considered.

Maintaining in memory estimates for every state $s$, or for every state-action pair $s,a$, can be very resource-intensive or just unfeasible when there are very many states. In these cases, the agent would approximate the value function using approximators with less parameters than states. Due to their high representational power, DNNs are nowadays widely used as approximators for the value functions in RL. All algorithms used in this paper use function approximation via a DNN.

\subsection{N-Dimensional Navigation}
From the considerations described at the beginning of this section, the problem of generating PRNs seems suitable for RL. There is however one caveat: the naive approach that comes to mind, that is, using as state the last generated number, is inherently not Markovian: whatever reward we use to measure the randomness of the sequence, it must depends on the whole sequence that was generated before. On the other hand, using as state the whole sequence, and increasing the length of the sequence by appending a new number is calling for the curse of dimensionality!

We address this issue by a completely different approach. We keep the sequence length fixed, say $N$. At each time step the agent can now fully observe the state of the environment, that is, the entire sequence of length $N$, and this state is Markovian by definition. In this paper, we consider two different formulations, one where the state is a sequence of decimal numbers and another where the state is a sequence of bits.

The goal of this decision task is to perturb the sequence without changing its length, and this can be obtained by adding or subtracting a certain fixed value at some position in the sequence. Actions will then be given by "add or subtract, position" pairs, and will be described more in detail in next subsections concerning each specific formulation.

Finally, we model the task as finite-horizon episodic, by fixing a termination time $T$. Thus, the state $S_{T}$ is the output sequence of the PRNG, i.e.\ the sequence of $N$ PRNs. A fixed length output however violates one requirement of PRNGs, that is, a PRNG has to be able to generate a (possibly) infinite amount of numbers. To solve this problem, we can think of the generated sequence as the period of the PRNG. By concatenating multiple output of the same PRNG it is possible to obtain a (possibly) infinite amount of numbers. Please note that this will produce PRNs with ``variable period'', because the learned policy will generally be stochastic, as we will show in the next section. This is a feature of this RL approach.

As usual, we start generating PRNs from a number known as the seed. According to our definition of state, we set all $N$ numbers of the sequence corresponding to the starting state equal to the seed value, e.g.\ 0.

Both decimal and binary formulation are in general an $N$-dimensional navigation task. To show how this is true, suppose $N = 3$ and a seed $s_0=0$.
Moreover, assume that each number belongs to the range $\{-10, \dots , 10\}\subset\mathbb{Z}$, where each element is represented in integer decimal form. We define a set of actions
\[
\mathcal{A}:=\{+1_0, -1_0, +1_1, -1_1, +1_2, -1_2, \pass\}
\]
where $+1_0$ means to add $+1$ to the number at position $0$, $-1_0$ means to add $-1$ to the number at position $0$ and so on. The action $\pass$ means to do nothing, for a total of $7$ actions. The seed $s_0$ gives the starting state
\[
S_0 = [0, 0, 0].
\]
Now, at the initial time step $t = 0$ the agent observes $S_0$ and takes the action $+1_2$. As a consequence, the environment goes in the state
\[
S_1 = [0, 0, 1].
\]
Now, the cycle starts over and at the next time step $t = 1$ the agent observes $S_1$ and takes the action $-1_0$. As a consequence, the environment answers with the state
\[
S_2 = [-1, 0, 1].
\]
At time step $t=T-1$, the agent takes the last action, because after that the episode ends. Suppose the terminal state is
\[
S_{T} = [-7, 3, 10].
\]
We can geometrically describe the above agent-environment interaction as a navigation problem in $\mathbb{Z}^3$: we started from the origin $(0,0,0)$, and moved $T$ times along axis, ending up in $(-7,3,10)\in\mathbb{Z}^3$. When actions are chosen by a policy maximizing a total reward associated to ``randomness'' in some sense, this navigation will end up in a PRNs sequence.

A simple yet informative reward function could be the average of the values given by the NIST tests computed over the binary representation of the sequence at certain time steps. The NIST test suite and its tests are described more in detail in section~\ref{sec:nist}.

With a sequence of arbitrary length $N$, the problem scales to an $N$-dimensional navigation problem.

\subsection{Decimal Formulation}
The first proposed formulation is based on decimal integer numbers and is called Decimal Formulation (DF). This formulation introduces a strong bias while trying to keep the sequence length $N$ and the set of actions as small as possible. We believe it is also the most intuitive formulation and as such is presented first. Given a sequence of length $N$, the state is the $N$-dimensional list
\[
S = [x_1, x_2, \dots, x_N]
\] 
where $x_1$, $x_2$, \dots, $x_N$ are integer values defined inside a certain range. Usually, the range is at most determined by the binary representation supported by the NIST test suite implementation. If an $8$-bits representation is used values are defined as 
\[
x\in\{-128, \dots, 127\}
\] 
where $x$ is any integer decimal number belonging to the sequence. The action set $A$ is defined as
\[
\mathcal{A} = \bigcup_{n=1}^N\{+1_n, -1_n, \pass\}
\] 
where $+1_n$ is the action of adding $+1$ at the $n$-index of the sequence, $-1_n$ is the action of adding $-1$ at the $n$-index of the sequence and $\pass$ is the null-action, i.e.\ do nothing. Adding $+1$ at $S=128$, or $-1$ at $S=-127$, is equivalent to a \pass. The action set is discrete and its size is $2*N+1$. DF, while intuitive, showed a major caveat: episodes frequently ends nearby the starting state. This is due to the sparsity of good states in $\mathbb{Z}^n$: until the RL agent doesn't experience a good state, it cannot improve its policy, and the generated sequence on the long term will be determined by the starting policy, usually uniform. 

\subsection{Binary Formulation}
Binary Formulation (BF) is based directly on the binary representation of the sequences as analyzed by the NIST test suite. It overcomes the correlation between starting and terminal states of the DF, probably because in this case good states are less sparse. Assume each integer value is represented by $8$-bits. Given a sequence of $N$ integers, the state is
\[
S = [b_1, b_2, \dots, b_8, b_9, \dots, b_{8N}]
\]
where $b_1$, $b_2$, \dots, $b_8$, $b_9$, \dots, $b_{8N}$ are the binary values of the $N$ integers. More in general, given a $m$-bit binary representation, we have a state space of cardinality $2^B = 2^{m\cdot N}$. The action set is defined as 
\[
\mathcal{A} = \bigcup_{n=1}^{B}\{1_n, 0_n\}
\] 
where $1_n$ is the action of setting the $n$ bit to $1$ and $0_n$ is the action of setting the $n$ bit to $0$. The null action is not required since to keep current position in the space the agent can just set a certain bit to its current value. The action set is discrete and its size is $2\cdot B$, i.e.\ $2\cdot m\cdot N$. While binary representation can produce terminal states uncorrelated to starting states, the size of the action set is large even for short decimal sequences.

\subsection{NIST Test Suite}\label{sec:nist}
The NIST statistical test suite for random and pseudo-random number generators is the most popular application to test the randomness of sequences of bits. It has been published as a result of a comprehensive theoretical and experimental analysis and may be considered as the state-of-the-art in randomness testing for cryptographic and not cryptographic applications. The test suite has become a standard stage in assessing the outcome of PRNGs shortly after its publication. 
 
The NIST test suite is based on statistical hypothesis testing and contains a set of statistical tests specially designed to assess different PRNs sequences properties. An hypothesis test is a procedure for determining if a given assertion is true. In this case the provided \emph{P-values} determines whether or not the tested sequence is random from the perspective of the selected statistic. This value could be thought as the value returned by the test. Each statistical test is designed to check for a relevant randomness statistic and is formulated to test a null hypothesis $H0$ on it. The null hypothesis under test is that the sequence being tested is random from the point of view of the current selected statistic, and the alternative hypothesis $Ha$ is that the tested sequence is not random. Mathematical methods determine a reference distribution of the selected statistic under the null hypothesis and a critical value is accordingly defined. Each test then derives a decision based on the comparison between the critical value and the test statistic value computed on the sequence. If the test value is less than the critical value the test is said to be rejected, otherwise is said to be passed. According to this decision the test accepts or rejects the null hypothesis and concludes whether the tested generator is or is not producing PRNs w.r.t.\ the selected statistic. 
 
The NIST test suite is composed by the following tests:
\begin{itemize}
    \item \textbf{The Frequency (Monobit) Test}: this frequency test determines whether zero and one bits appear in the tested sequence with approximately the same probability. This simple test can then reveal the most obvious deviations from randomness.
    \item \textbf{Frequency Test within a Block}: this frequency test is a generalization of the previous Frequency (Monobit) Test, having the purpose of determining the frequency of zeroes and ones within $M$-bit blocks and thus revealing whether zeroes and ones are uniformly distributed throughout the tested sequence. 
    \item \textbf{Runs Test}: in order to determine whether transitions between zeroes and ones in the sequence appear as often as expected from a random sequence, this test counts the total number of \emph{runs} of various lengths. A run consists of an uninterrupted sequence of identical bits.
    \item \textbf{Longest Run of Ones in a Block Test}: in this test, the sequence is processed in $M$-bit blocks with the aim of determining whether the length of the longest run of ones in a block is consistent with the length expected from a random sequence. 
    \item \textbf{Binary Matrix Rank Test}: the focus of this test is to compute the rank of disjoint sub-matrices of the entire sequence. Its purpose is to check for linear dependence among fixed length substrings of the original sequence. 
    \item \textbf{Discrete Fourier Transform (Spectral) Test}: the focus of this test is to count the peak heights in the Discrete Fourier Transform of the sequence. Its purpose is to detect periodic features (i.e., repetitive patterns that are near each other) in the tested sequence that would indicate a deviation from the assumption of randomness. 
    \item \textbf{Non-Overlapping Template Matching Test}: the focus of this test is the number of occurrences of pre-specified target strings. Its purpose is to detect generators that produce too many occurrences of a given non-periodic (aperiodic) pattern. 
    \item \textbf{Overlapping Template Matching Test}: this test is similar to the Non-Overlapping Template Matching Test, but it extends the search criteria to overlapping patterns.
    \item \textbf{Maurer’s ``Universal Statistical'' Test}: the focus of this test is the number of bits between matching patterns (a measure that is related to the length of a compressed sequence). Its purpose is to detect whether or not the sequence can be significantly compressed without loss of information.
    \item \textbf{Linear Complexity Test}: the purpose of this test is to determine the linear complexity of the \emph{LFSR (Linear Feedback Shift Register)} that could generate the tested sequence. If the complexity is not sufficiently high, the sequence is non-random. 
    \item \textbf{Serial Test}: in order to verify the uniformity of templates the test counts the occurrences of every possible $M$-bit overlapping patterns in the sequence. A high level of uniformity–patterns occur with the same probability indicates that the sequence is close to random.
    \item \textbf{Approximate Entropy Test}: the purpose of this test is to compare the frequency of overlapping patterns of two consecutive lengths against the expected frequency of a true random sequence.
    \item \textbf{Cumulative Sums (Cusum) Test}: the focus of this test is the maximal excursion (from zero) of the random walk defined by the cumulative sum of adjusted $(-1, +1)$ digits in the sequence. Its purpose is to determine whether the cumulative sum of the partial sequences occurring in the tested sequence is too large or too small relative to the expected behavior of that cumulative sum for random sequences.
    \item \textbf{Random Excursions Test}: the focus of this test is the number of cycles having exactly $K$ visits in a cumulative sum random walk. The cumulative sum random walk is derived from partial sums after the $(0,1)$ sequence is transferred to the appropriate $(-1, +1)$ sequence. Its purpose is to determine if the number of visits to a particular state within a cycle deviates from what one would expect for a random sequence.
    \item \textbf{Random Excursions Variant Test}: the focus of this test is the total number of times that a particular state is visited (i.e., occurs) in a cumulative sum random walk. Its purpose is to detect deviations from the expected number of visits to various states in the random walk. 
\end{itemize} 

In this paper, the NIST test suite is used to compute the average value of all eligible tests in the battery, run on the generated PRNs sequences. Some tests return multiple statistic values: in that case, their average is taken. If a test is failed its value in the average is set to zero. Some tests are not eligible on certain sequences because of their length, and in this case they are not considered for the average.

This value is then used as a reward function for the MDP we are defining. Since this is an exploratory paper, we decided to assign rewards in two opposite ways: at every time step, and only at the end of the episode. The first choice has the effect of maximizing an expected randomness during trajectories, and was made with the idea that encouraging the agent to find good states with good nearby states could result in a more stable training. Of course, this could lead to the undesirable effect of finding a low value state at the end of an episode where all the preceding states have high values.
The second choice was made with the (in some sense, opposite) idea that we are more interested in a terminal state with a very high value, rather than in a trajectory full of states with decent values. This seems the most clean way to reach the goal.
However, this environment can be harder for agents to train on: since the reward is assigned only at the end, it will affect weakly all temporally distant states that have preceded it. This is known as the credit assignment problem, and usually leads to bad convergence properties of the RL algorithms whenever the average length of the episodes is large enough.

Please notice that, since test statistic values are probabilities according to statistical hypothesis definition, rewards belong to $[0, 1]$.

\section{Experiments}
Our experiments consists on multiple sets of training processes with different hyperparameters. The goal of these experiments is not to achieve the best result given the setting, but instead to assess the feasibility of the setting and to infer which algorithms are better suited to solve the task.

We divide the experiments according to formulation (DF or BF), whether the rewards are given at each step or only at the end of the episodes, and by the RL algorithm employed. The combination of the first two gives the RL environment, while the latter is the RL agent. 

We experimented with three model-free RL algorithms: two on-policy policy gradient algorithms (Vanilla Policy Gradient and Proximal Policy Optimization) and one off-policy (deep Q-learning in its Dueling Deep Q-Network flavor). Details about these agents are given inside their relative subsections, and plots of their performances are instead analyzed in the last subsection, as well as general results.

All the tasks are episodic, with a fixed termination time $T$.

\subsection{Environments}
The environments used in our experiments are the following:
\begin{itemize}
    \item DF, rewards at every step. This environment uses DF for states and actions, and assigns a reward $R_t$ at every time step $t$. The reward is the average value of the NIST test suite over the binary representation $B_t$ of the state $S_t$ at the current time step $t$:
    \[
    R_t = \avgnist(B_t),\qquad t \in \{0, \dots, T\}.
    \]
    \item BF, rewards at every step. This environment uses BF for states and actions, and assigns a reward $R_t$ at every time step $t$. The reward is the average value of the NIST test suite over the state $S_t$ at the current time step $t$:
    \[
    R_t = \avgnist(S_t),\qquad t \in \{0, \dots, T\}.
    \] 
    \item BF, reward only at the end. This environment uses BF for states and actions, and assigns a reward $R_t=0$ at every time step $t\neq T$. At termination time $T$ the reward is the average value of the NIST test suite over the state $S_T$:
    \[
    R_t =
    \begin{cases}
        \avgnist(S_t) & \text{ if } t = T \\
        0               & \text{ otherwise}
    \end{cases}
    \] 
\end{itemize} 

\subsection{Framework}
The framework used for the RL algorithms is USienaRL\footnote{Available on PyPi and also on GitHub: https://github.com/InsaneMonster/USienaRL.}. This framework allows for environment, agent and interface definition using a preset of customizable models. While agents and environments are direct implementations of what is described in the RL theory, interfaces are specific to this implementation. Under this framework, an interface is a system used to convert environment states to agent observations, and to encode agent actions into the environment. This allows to define agents operating on different spaces while keeping the same environment. By default an interface is defined as pass-through, i.e.\ a fully observable state where agents action have direct effect on the environment. 

The NIST test battery is run with another framework, called NistRng\footnote{Available on PyPi and also on GitHub: https://github.com/InsaneMonster/NistRng.}. With this framework is possible to run a customizable battery of statistical set over a certain sequence. The framework also computes which tests are available over certain sequences, i.e.\ due to their length. Available tests for a certain sequence are called eligible tests. Each test returns a value and a flag stating if the test was successfully exceeded or not by the sequence. If a test is not eligible with respect to a certain sequence it cannot be run and it is skipped.

The code for this article can be found at this GitHub repository: \emph{https://github.com/InsaneMonster/pasqualini2019prngrl}

\subsection{Deep Q-Learning}
Deep Q-learning is the function approximation version of Q-learning. The policy iteration step in Q-learning chooses the next action in an exploratory fashion, and uses a mixed update rule with a greedy target. Accordingly, denoting by $Q(s, a; \theta)$ the DNN approximating $q$-values, its vector of parameters $\theta_t$ at time $t$ is updated towards the target $R_{t+1}+\max_{a\in\mathcal{A}}Q(S_{t+1}, a; \theta_t)$, where the reward $R_{t+1}$ is chosen by an exploration policy. In our experiments, we used as exploration policy a softmax distribution over $Q(s, a; \theta)$ with temperature $\tau$ (this is called Boltzmann sampling, and it is slightly more efficient than the $\epsilon$-greedy exploration policy used in the original paper \cite{atari}).
Temperature is reduced over the course of training.
Q-Learning algorithms which use DNNs as $q$-value approximators are called Deep Q-Networks (DQN) algorithms. In this paper, we use the Dueling Double DQN \cite{wang2015dueling}, a DQN flavour which is reported as overall more stable and on average is proved to learn better than the original DQN algorithm from \cite{atari}.
Finally, we use Prioritized Experience Replay as in \cite{schaul2015prioritized}.

\subsection{Vanilla Policy Gradient}
Vanilla Policy Gradient (VPG) \cite{sutton2000policy} is an on-policy RL algorithm whose aim is to learn a policy without using $q$-values as a proxy. This is obtained increasing the probabilities of actions that lead to higher return, and decreasing the probabilities of actions that lead to lower return.
The agent at each time step $t$ uses a DNN to predict a probability distribution over its actions space $\pi(a|s; \theta)$, where $\theta$ are the DNN parameters.
VPG works by updating policy parameters via stochastic gradient ascent on policy performance over a buffer built from a certain number of trajectories: 
\[
\theta \longleftarrow \theta + \alpha \nabla_{\theta} J(\pi(\cdot|\cdot; \theta))
\] 
where $J(\pi(\cdot|\cdot; \theta))$ denote the expected finite-horizon undiscounted return of the policy and $\nabla_{\theta}$ its gradient with respect to $\theta$. To compute $J(\pi(\cdot|\cdot; \theta))$ a computation of advantages $\hat{A_t}$ for each state $s$ in the buffer is required. In this paper we use Generalized Advantage Estimation (GAE) \cite{schulman2015high} to compute $\hat{A_t}$, and saved rewards $R_t$ are normalized with respect to when they are collected (the so called rewards-to-go). These are solutions reported in literature to be stable and to improve overall training performance of the model.

\subsection{Proximal Policy Optimization}
Proximal Policy Optimization (PPO) \cite{schulman2017proximal} is another on-policy RL algorithm which improves upon VPG. PPO tries to take the biggest possible improvement step on a policy using the data it currently has, without stepping too far and making the performance collapse. This algorithm is considered state-of-the-art in policy optimization methods. In the implementation used in this paper, when parameters $\theta$ are updated over a buffer of trajectories, a certain value of approximated KL divergence between old policy and new policy is computed. If this value is larger than a certain threshold $K_{th}$, update is early stopped. Similarly to VPG, computation of advantages $\hat{A_t}$ for each state $s$ in the buffer is required. Just like with the previous on-policy algorithm, GAE is used to compute $\hat{A_t}$ and saved rewards $R_t$ are normalized as rewards-to-go.

\subsection{Results}
In this subsection we present experimental results in the form of plots.

First we analyze the training process of Dueling Double DQN and VPG algorithms over DF, reward at every step environment. To measure the quality of training we measure the \emph{average scaled reward} over last 20 steps of each episode for each volley. By scaled reward we mean the average reward per step. The average scaled reward over the volley is then the average, over the episodes belonging to the volley, of the average reward per step. The length of the sequence is $N = 10$ so the task is a $10$-dimensional navigation task.

The Dueling Double DQN model used in these experiments is composed by six dense layers with 2048 neurons each initialized with Xavier initialization. The learning rate is set to $1e-6$ with discount factor $\gamma = 0.6$. Weights of the Q-Network are copied to the target network every 1000 steps. The batch size is 150 and the exploration policy during training is Boltzmann sampling with temperature decay of $2e-6$ per episode. The VPG model used in these experiments is composed by six dense layers with 2048 neurons each initialized with Xavier initialization, the learning rate of the policy stream is set to $3e-5$ and the learning rate of the value stream is set to $1e-4$. Discount factor is $\gamma = 0.6$ and $\lambda = 0.95$. At each update, 80 value steps are performed. The model is updated 20 times each volley. 

\begin{figure}[!tbp]
  \centering
  \subfloat[Dueling Double DQN]{\includegraphics[width=0.5\textwidth]{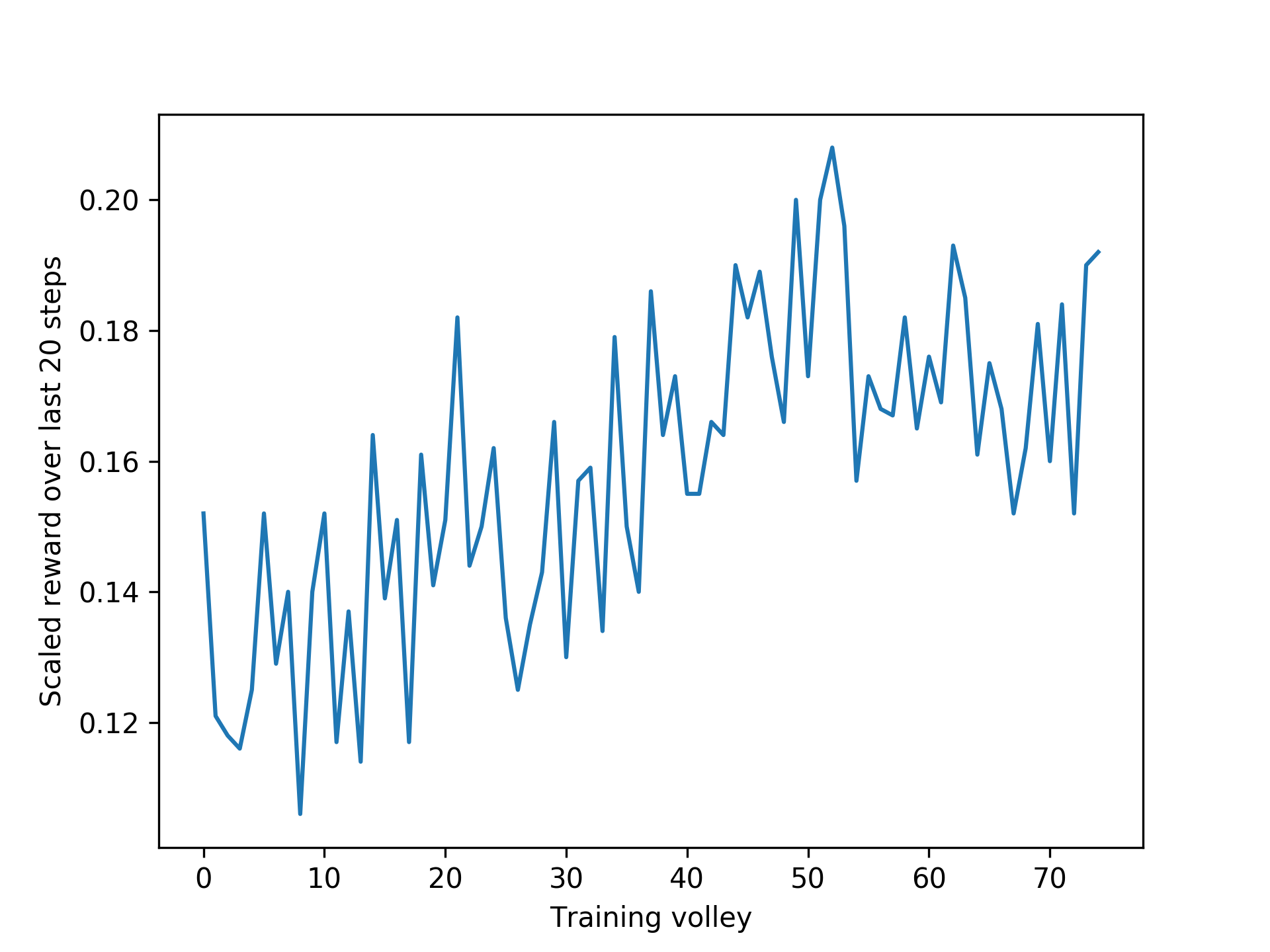}\label{fig:dfna}}
  \hfill
  \subfloat[Vanilla Policy Gradient]{\includegraphics[width=0.5\textwidth]{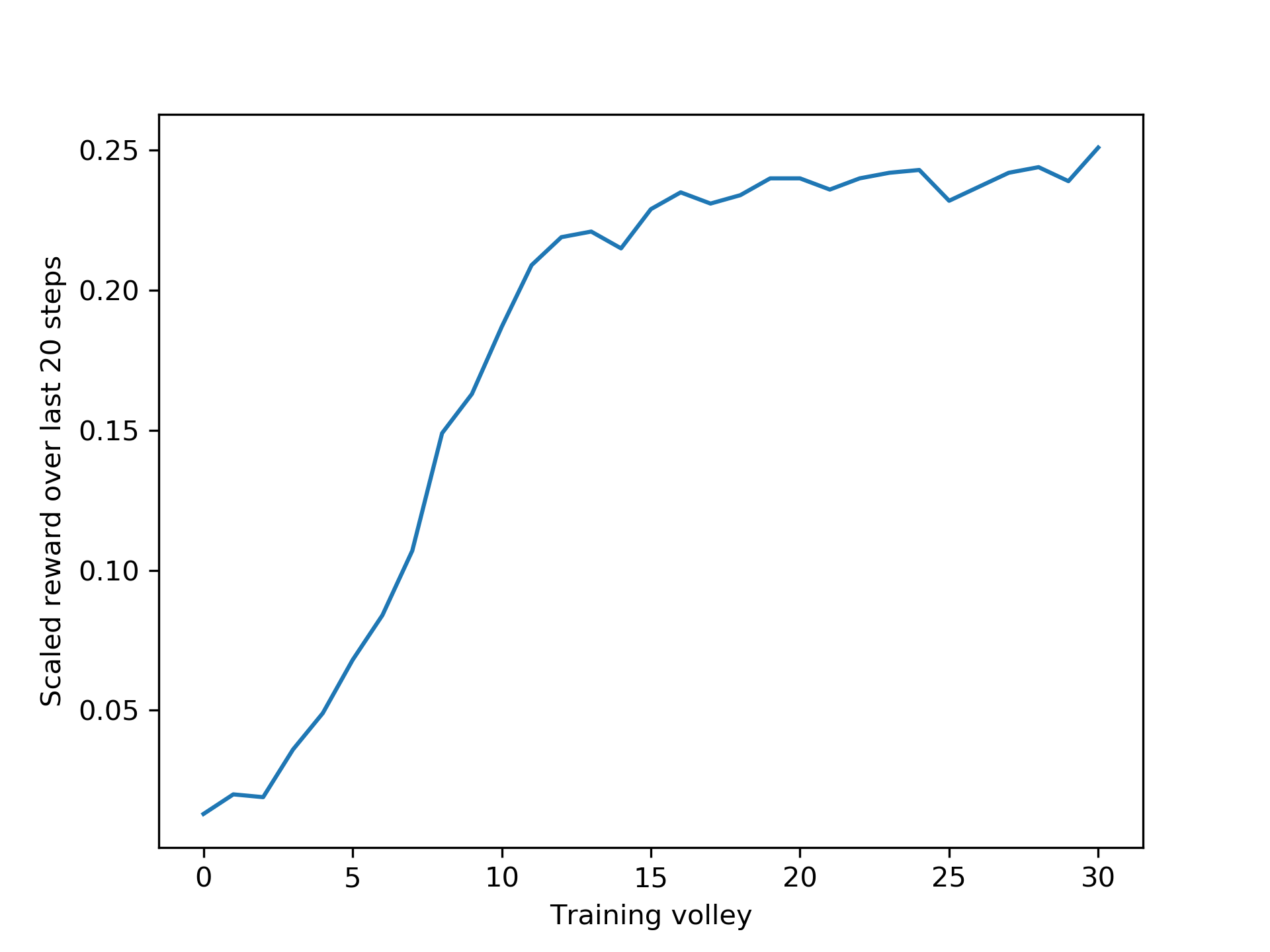}\label{fig:dfnb}}
  \caption{average scaled rewards over last 20 steps on decimal not-episodic environments during training of Dueling Double DQN (\ref{fig:dfna}) and of VPG (\ref{fig:dfnb}). Volleys are composed by 1000 episodes each. The fixed length of each trajectory is $T = 30$ steps.}
  \label{fig:dfn}
\end{figure}

We can see in figure~\ref{fig:dfna} that the trend, while positive, is very unstable. This is to be expected since it is very hard for Q-Learning to converge in a problem with such a large action space and such complex reward function. As consequence, Q-Learning and its variants seem to be not suitable for completing the task at hand. In figure~\ref{fig:dfnb} we can see that the trend with VPG is much more stable and overall converges to better solutions, faster.

Then we analyze the BF, reward at every step environment. Here we use as length of the sequence the number of bits belonging to the $8$-bit binary representation of the sequence. We set $B = 80$ so it's a sequence of 80 bits and the task is a $80$-dimensional navigation task. Experimentally we saw the action space to be too large to handle for Dueling Double DQN, so we only run VPG algorithm.

The VPG model used in these experiments is the same as the one defined in the DF, reward at every step experiments, just with doubled neurons for each layer. Like in the previous experiments, we use as criteria the average scaled reward over each volley. 

\begin{figure}
    \centering
    \includegraphics[width=0.65\textwidth]{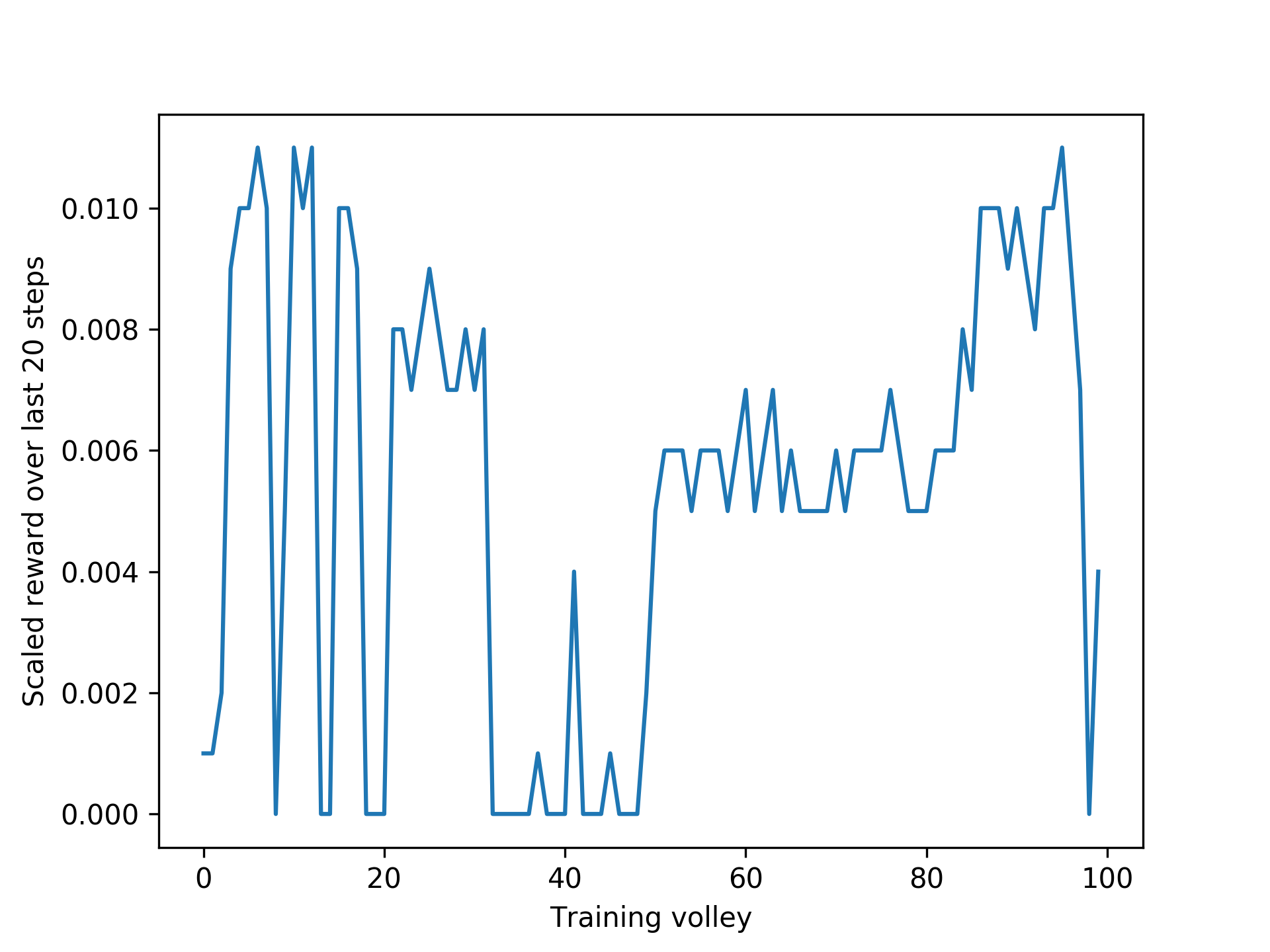}
    \caption{average scaled rewards over last 20 steps on binary not-episodic environments during training of VPG. Volleys are composed by 1000 episodes each. The fixed length of each trajectory is $T = 20$ steps.}
    \label{fig:bfn}
\end{figure}

We can see in figure~\ref{fig:bfn} that the system tends to be very unstable and overall not reaching satisfying results.

{}

Last we train our models over the BF, reward only at the end environment, with $B = 80$ and $B = 200$.
We first test VPG algorithm and then the PPO algorithm. To measure the quality of training we measure the average reward per episode over each volley.

The VPG model used in these experiments is composed by six dense layers with 4096 neurons each initialized with Xavier initialization, the learning rate of the policy stream is set to $3e-4$ and the learning rate of the value stream is set to $1e-4$. Discount factor is $\gamma = 0.99$ and $\lambda = 0.95$. At each update, 80 value steps are performed. The model is updated 10 times each volley. The PPO model is likewise composed by six dense layers with 4096 neurons each initialized with Xavier initialization. The learning rate of the policy stream is set to $3e-4$ and the learning rate of the value stream is set to $1e-4$. Discount factor is $\gamma = 0.99$ and $\lambda = 0.97$. At each update, 80 policy and value steps are performed. The model is updated 10 times each volley. The clip ratio is set to $0.2$ and the target KL divergence is $0.01$.

\begin{figure}[!tbp]
  \centering
  \subfloat[Vanilla Policy Gradient with $B = 80$]{\includegraphics[width=0.5\textwidth]{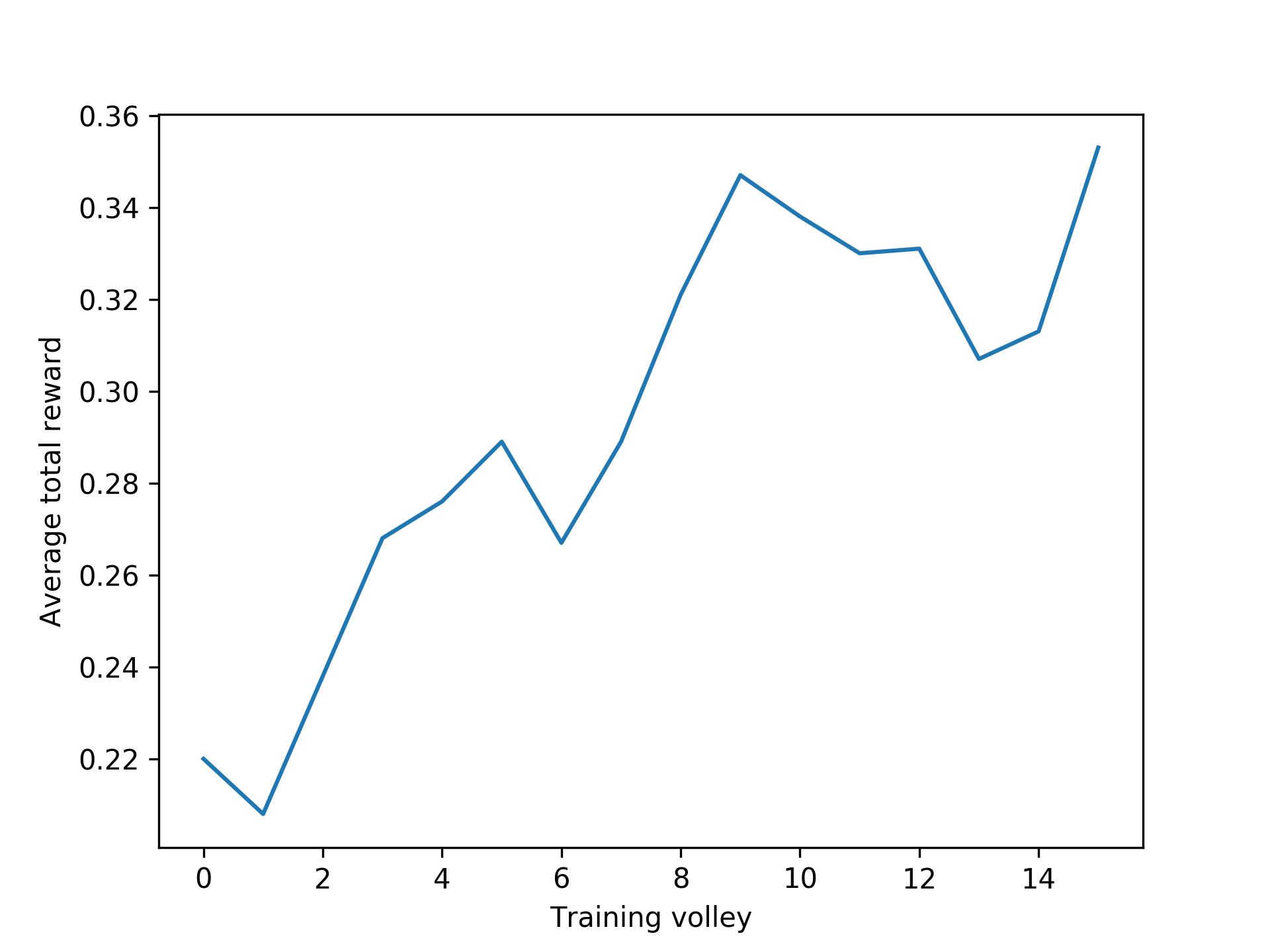}\label{fig:bfe80a}}
  \hfill
  \subfloat[Proximal Policy Optimization with $B = 80$]{\includegraphics[width=0.5\textwidth]{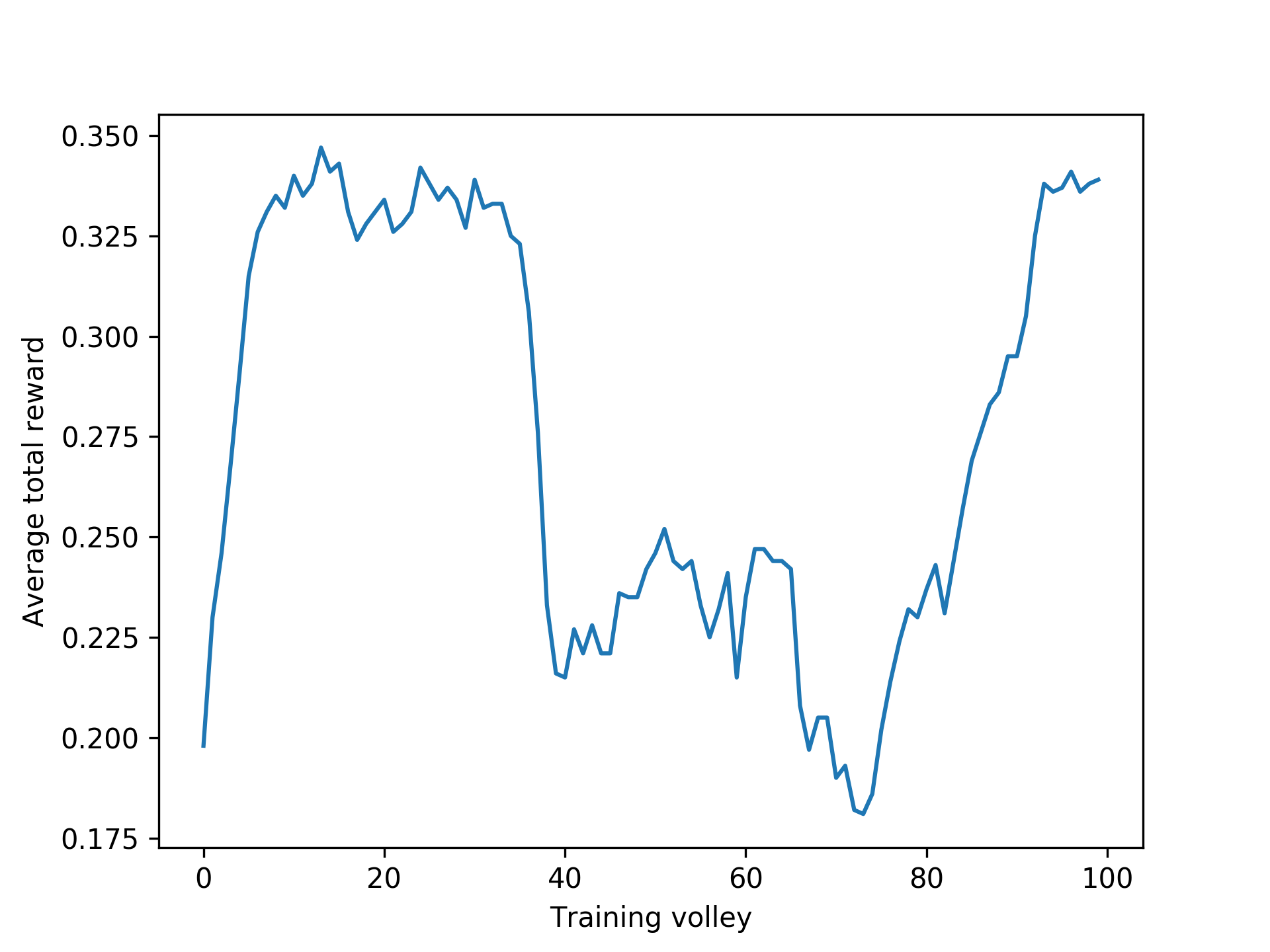}\label{fig:bfe80b}}
  \caption{Average total reward during training of VPG (\ref{fig:bfe80a}) and PPO (\ref{fig:bfe80b}) on BF, reward only at the end environment with $B = 80$. Volleys are composed by 1000 episodes each. The fixed length of each trajectory is $T = 100$ steps. Reference average NIST value over 1000 sequences of same binary length $B = 80$ generated by NumPy uniform PRNG is 0.33.}
  \label{fig:bfe80}
\end{figure}

\begin{figure}[!tbp]
  \centering
  \subfloat[Vanilla Policy Gradient with $B = 200$]{\includegraphics[width=0.5\textwidth]{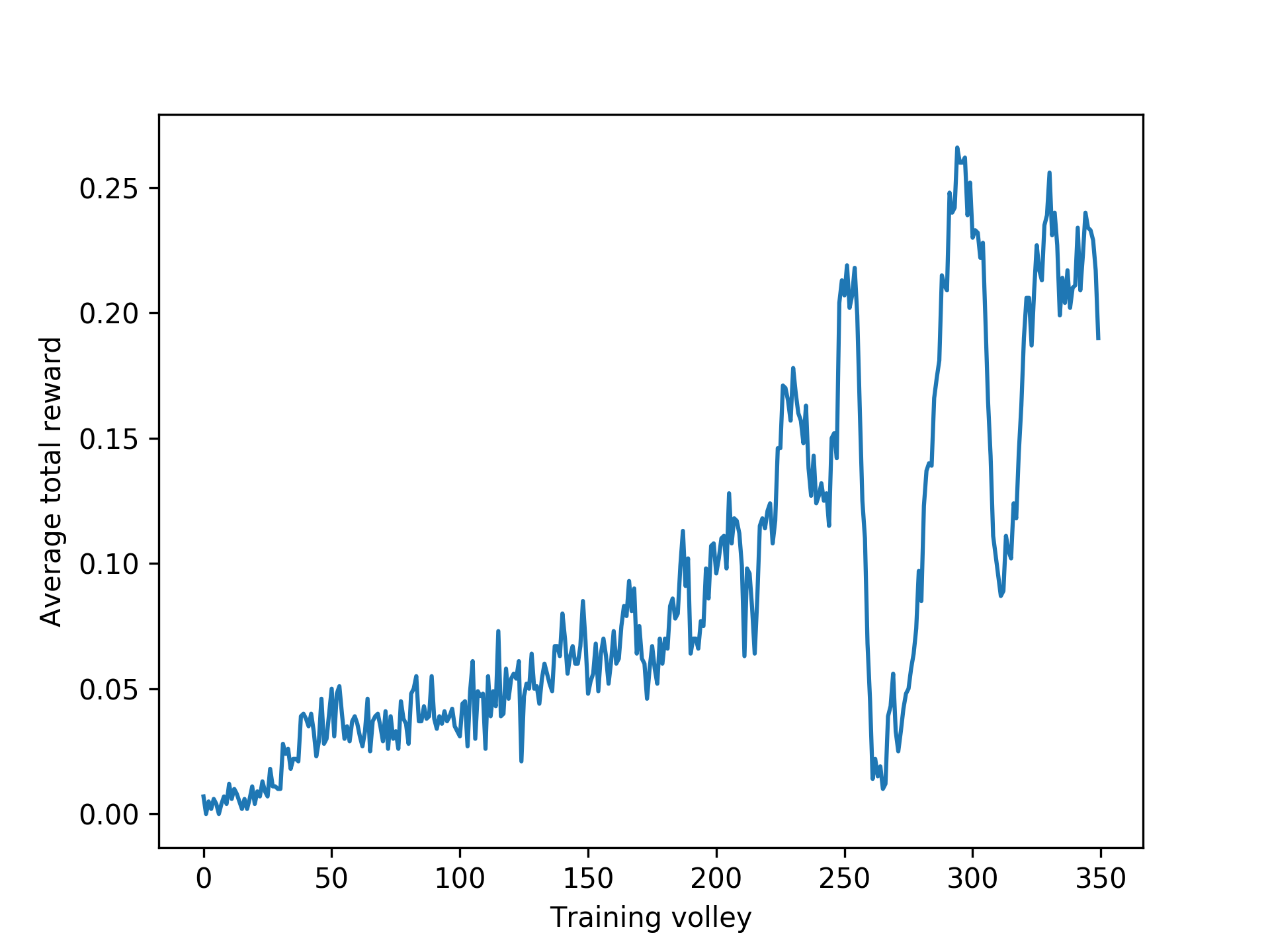}\label{fig:bfe200a}}
  \hfill
  \subfloat[Proximal Policy Optimization with $B = 200$]{\includegraphics[width=0.5\textwidth]{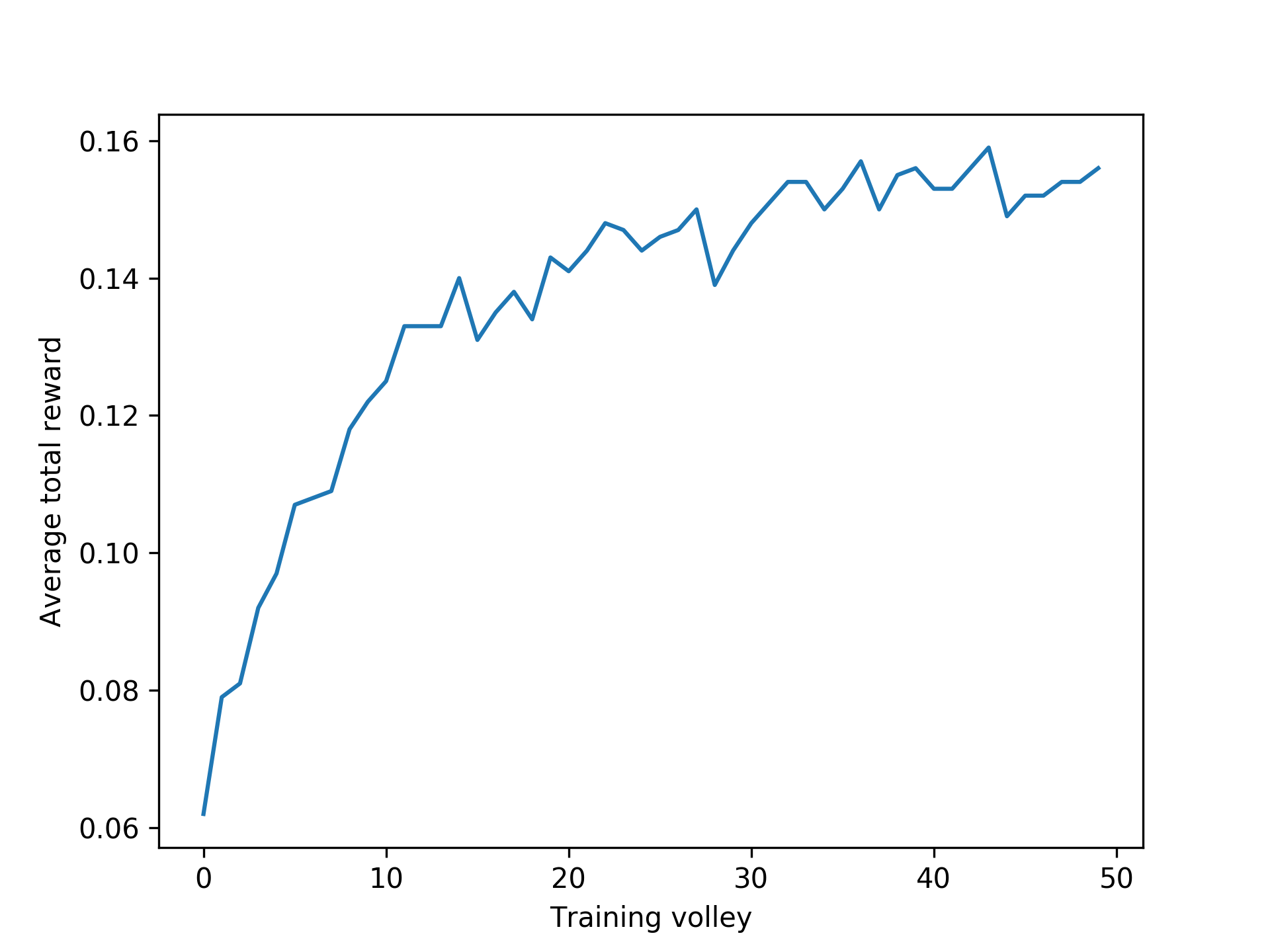}\label{fig:bfe200b}}
  \caption{Average total reward during training of VPG (\ref{fig:bfe200a}) and PPO (\ref{fig:bfe200b}) on BF, reward only at the end environment with $B = 200$. Volleys are composed by 1000 episodes each. The fixed length of each trajectory is $T = 100$ steps. Reference average NIST value over 1000 sequences of same binary length $B = 200$ generated by NumPy uniform PRNG is 0.35.}
  \label{fig:bfe200}
\end{figure}

We can see in figure~\ref{fig:bfe80} that for sequences of 80 bits both algorithms converge with consistent results. We also note that our model average value is greater than the reference average value over 1000 sequences of equal fixed length $B = 80$ generated by the NumPy uniform PRNG. We believe this to be a very interesting result. Apparently, for sequences of length 200 VPG is able to converge, as seen in~\ref{fig:bfe200a}, but not in all experiments. PPO, while having better training trends overall as shown in~\ref{fig:bfe200b}, proves to be even too cautious in improving its policy and tends to converge to worse final results or to converge in more steps. In table~\ref{tab:seq} some generated sequences are shown in decimal representation. All the generated sequences are results of one trained PRNG with one seed state. Since any fixed length sequence can be considered the period of the generator and there are multiple sequences given by just one generator, the trained PRNG has variable period.

\begin{table}
    \centering
    \begin{tabular}{lllllllllllllll}
        \toprule
        \cmidrule(r){1-14}
        \multicolumn{11}{c}{Sequence} & \multicolumn{4}{c}{Average NIST Value}    \\
        \midrule
        -10 & -112 & 68 & -39 & -123 & 35 & -45 & 66 & -28 & 62 & & & & 0.24 \\
        22 & -113 & 34 & -111 & 44 & 42 & 63 & 114 & -63 & -41 & & & & 0.57  \\
        -48 & -111 & 20 & -102 & 10 & -18 & 55 & 11 & 80 & 62 & & & & 0.16   \\
        \bottomrule
    \end{tabular}
    \vspace*{+3mm}
    \caption{Sequences in decimal representation of one trained PRNG with $seed = 0$ and $B = 80$, alongside their average NIST value.}
    \label{tab:seq}
\end{table}

\section{Conclusions}
In this paper we propose a way to automatically generate PRNGs, a task of interest and a currently open field of research. Our approach uses RL to build a PRNG from scratch. To the best of our knowledge, this is a novel approach.

Results are promising, in particular when sequences are represented as sequences of bits and the reward is assigned only at the end of episodes.
Our approach also presents the following interesting features:
\begin{itemize}
    \item It requires no input data, so that the generated PRNG is always a novel algorithm, and the trained agent explores solutions possibly unknown to human-generated or supervised-learning-generated algorithm. In other words, the generated PRNG is not an imitation of some other algorithm and the generated sequence is not an imitation of a pre-existing PRNs or TRNs sequence. 
    \item Each time a training process is run, the resulting PRNGs are likely to be different from each other. This is an inherent property of RL as a whole: the policy learned is one of the very many stochastic optimal policies. Moreover, one can change the model hyperparameters and/or the training algorithm to increase diversity between generated PRNGs.
    \item For a single starting state, multiple solutions can be obtained after one training process since RL policies can be stochastic. Indeed, given a certain starting state, the same agent will usually follow different trajectories, leading to different output sequences. In short, we obtain a non-deterministic PRNG given a single seed. This is a novel property which is not present in current state-of-the-art PRNGs.
    \item Given that NNs are black-box approximators, the policy of the RL agent is black-box. Since the PRNG is the algorithm given by that policy, it is also black-box. This grants the nice property of having no human insights in the inner functioning of the PRNG, thus providing (potentially) increased security from a cryptographic standpoint.
\end{itemize}

Finally, results prove that this approach is feasible, and the task can be learned by RL techniques. We hope this paper will inspire future research which combines PRNGs and RL.

\subsection{Future work}
The current main limitation of our approach is the dimensionality $N$ of the state, i.e.\ the period of the PRNG. Our experiments show that, at least on average hardware, is very complex to successfully train an agent on longer sequences, thus obtaining a PRNG with longer period.
This will be the main focus of our research to come. Beside that, we aim to do the following: 
\begin{itemize}
    \item Improve the quality of the output sequence, i.e.\ each period, especially for longer sequences. The more uniformly distributed (i.e.\ similar to a TRNs sequence) the bits are, the more an accurate result is also obtained by concatenating multiple sequences. This means to increase the average value obtained by the NIST test battery over the fixed size sequences.
    \item Increase the amount of supported seeds. We aim at reducing the variance introduced by additional seeds during learning. We believe that processing somehow the vector representation (for example with convolutional filters) could be a promising path to follow.
    \item Devise a better strategy to concatenate the output sequences. Ideally, this could also be learned, for example with hierarchical RL.
    \item Move towards a formulation in which the size of the action space does not grow (too much) with the length $N$ of the sequence, without losing (too much) the ability to reach points in the lattice with a high value.
\end{itemize}
\bibliographystyle{splncs04}
\bibliography{references}

\end{document}